\documentclass{aastex}
\usepackage{emulateapj5}
\usepackage{apjfonts}

\newcommand{\ea}{{\it et al.}}

\newcommand{\msol}{\mathrm{M}_\odot}

\newcommand{\kms}{km~$\rm{s}^{-1}$}

\newcommand{\beq}{\begin{equation}}
\newcommand{\eeq}{\end{equation}}
\newcommand{\bdm}{\begin{displaymath}}
\newcommand{\edm}{\end{displaymath}}

\slugcomment{Submitted to the Astrophysical Journal}

\begin{document}

\title{Strings in the $\eta$ Carinae nebula: hypersonic radiative cosmic bullets}

\author{A.Y. Poludnenko\altaffilmark{1,2}, A. Frank\altaffilmark{1,2},
S. Mitran\altaffilmark{3}}

\altaffiltext{1}{Department of Physics and Astronomy, University of Rochester,
Rochester, NY 14627-0171; wma@pas.rochester.edu, afrank@pas.rochester.edu}
\altaffiltext{2}{Laboratory for Laser Energetics, University of Rochester,
250 East River Road, Rochester, NY 14623}
\altaffiltext{3}{Department of Mathematics, University of North Carolina,
CB \#3250, Chapel Hill, NC 27599; mitran@amath.unc.edu}

\begin{abstract}
  We present the results of a numerical study focusing on the
  propagation of a hypersonic bullet subject to radiative cooling.
  Our goal is to explore the feasibility of such a model for the
  formation of ``strings'' observed in the $\eta$ Carinae Homunculus
  nebula. Our simulations were performed in cylindrical symmetry with
  the adaptive mesh refinement code AstroBEAR. The radiative cooling
  of the system was followed using the cooling curve by
  \citet{Dalgarno}. In this letter we discuss the evolution and
  overall morphology of the system as well as key kinematic
  properties. We find that radiative bullets can produce structures
  with properties similar to those of the $\eta$ Carinae strings, i.e.
  high length-to-width ratios and Hubble-type flows in the form of a
  linear velocity increase from the base of the wake to the bullet
  head.  These features, along with the appearance of periodic
  ``ring-like'' structures, may also make this model applicable to
  other astrophysical systems such as planetary nebulae, e.g. CRL 618
  and NGC 6543, young stellar objects, etc.

\end{abstract}


\section{INTRODUCTION}
Of the many intriguing questions surrounding the luminous blue
variable $\eta$ Carinae, the origin of the long thin ``strings'' found
in the outer nebular regions remains particularly vexing.  First
observed by \citet{Meaburn} and further studied by \citet*{Weis99} the
strings exhibit two remarkable properties: very high values of the
length-to-width ratios and the presence of the Hubble-type flow, i.e.
linear velocity increase from the base of the strings to their tip.
The strings also show a surface brightness decrease toward the string
head and the true tips possibly may be invisible in optical
wavelengths. These features represent the key challenges that must be
met by any model that hopes to offer a successful explanation of the
strings' origin and evolution.

Several models were suggested for the strings including jets
\citep{Garcia} and ionization shadows \citep{Soker}, see
\citet*{Redman} for a complete discussion of the suggested models. A
particularly promising model has been proposed by \citet{Redman} who
suggest that a single hypersonic bullet propagating in the ambient
medium is generating the strings observed in the $\eta$ Carinae
nebula. Since the Great Eruption that ejected the main nebula was know
to be an impulsive event with most of the mass and momentum directed
in the polar directions \citep{Smith}, and since the strings seem to
be coeval with the overall nebula, it is plausible that fragments or
bullets were produced in that same event.

In this work we do not address the issue of bullet origin focusing
instead on the nature of their evolution and observable properties. In
particular, we investigate the feasibility of the \citet{Redman} model
and we address two questions: (1) can a radiatively cooled hypersonic
bullet produce structures with large length-to-width ratios, and (2)
do such systems exhibit the Hubble-type flow behaviour in their
downstream wake.

In Section 2 we briefly discuss the numerical code used, the key
dimensionless parameters characterizing the system, and the setup of
the numerical simulation that was carried out. In Section 3 we discuss
the results of our numerical study, and in Section 4 we present the
conclusions including possible answers to the questions posed above as
well as applicability of our results to other astrophysical systems.

\section{NUMERICAL STUDY}

The simulation presented in this paper was performed with the
AstroBEAR code based on the BEARCLAW adaptive mesh refinement package
\citep{Mitran,Berger}. The code allows for arbitrary levels of
refinement within computer memory constraints, as well as adaptive
numerics and multiphysics. The Euler hydrodynamic equations with
cooling source terms were solved using explicit second-order--accurate
scheme.  The wave propagation method \citep{LeVeque97} was applied
with the full nonlinear Riemann solver. An operator split set of
ODE's, describing the source term effects, was solved using the
fourth-order--accurate fully implicit Kaps-Rentrop--type scheme
\citep{Kaps} (see also \citep{Recipes}). 


There are three dimensionless parameters that describe the evolution
of the system studied. The first two are the density contrast between
the bullet and the ambient medium $\chi_b \equiv \rho_b/\rho_a$ and
the Mach number $M_b$ of the bullet velocity at time $t \ = \ 0.0$.
Those two parameters determine the overall hydrodynamic regime of the
system evolution. The third one, the cooling parameter $\psi_b$,
describes the effects of cooling and is defined as 
\beq 
\psi_b =
\frac{t_{cool}}{t_{hydro}}. 
\label{coolpar} 
\eeq 
Here the hydrodynamic timescale $t_{hydro}$ is defined as the
bullet crushing time (or clump crushing time $t_{cc}$ as described by
\citet*{Pol})
\beq
t_{hydro} = \Big(\chi_b^{1/2}\big(F_{c1}F_{st}\big)^{-1/2}\Big)
\frac{2r_b}{v_{b}},
\label{thydro}
\eeq 
where $r_b$ is bullet radius, $v_b$ is bullet velocity, $F_{c1}
\approx 1.3$ and
\beq
F_{st} \simeq 1+\frac{2.16}{1+6.55\chi ^{-1/2}}.
\label{fst}
\eeq
The two factors $F_{c1}$ and $F_{st}$ relate the unperturbed
upstream conditions with the internal bullet post-shock ones and are
described in \citep{Pol}. The interaction of the bullet with the
ambient medium launches a shock propagating inside the bullet with the
post-shock temperature $T_{ps}$ and $\rho_{ps}$. The cooling rate
$\Lambda_{ps}$ for such a temperature can be found based on the
cooling curve given by \citet{Dalgarno}. Thus the cooling timescale is
estimated as follows
\beq
t_{cool}=\frac{T_{ps}-T_{min}}{(\gamma-1)\rho_{ps}\Lambda_{ps}}km_H,
\label{tcool}
\eeq
where $T_{min} = 100$ K is the minimum temperature gas can cool
down to. The internal bullet post-shock temperature $T_{ps}$ and
density $\rho_{ps}$ can be found in the usual manner using the
Rankine-Hugoniot relations. $T_{ps}$ and density $\rho_{ps}$ are the
functions of the unshocked bullet temperature and density respectively
as well as of the internal bullet shock Mach number $M_{is}$ described
by the expression
\beq
M_{is}=M_b\big(F_{c1}F_{st}\big)^{1/2}.
\label{Mis}
\eeq

We carried out simulations covering bullet Mach numbers in the range
$M_b \ = 10-200$. In terms of the cooling parameter we explored the
regimes ranging from the quasi-adiabatic one $\psi_b > 1$ to the
strongly cooling one $\psi_b \approx 10^{-5}$. In this paper we will
report only on the case that allows us to best address the issues
related to $\eta$ Carinae strings. From our parameter space
explorations this turns out to be the case with the strongest cooling
$\psi_b = 2.8\cdot10^{-5}$. We defer broader conclusions gleaned from
the parameter space exploration to a subsequent paper.

Here we present a simulation of bullet propagation where the initial
bullet radius is $r_b=3.0\cdot10^{15}$ cm. The initial bullet density
is $\rho_b=10^5 \ cm^{-3}$ in the inner 80\% of the bullet radius and
in the outer 20\% of the radius density is smoothed out through the
$tanh-$type function (cf. expression (1) in \citep{Pol}). This
translates into the initial bullet mass $m_b \approx 0.5\cdot10^{-5}
\msol$. The initial bullet Mach number is $M_b=20$, or $v_b \approx
235$ \kms. The computational domain is of size $1.8\cdot10^{17}$ cm
$\times \ 2.4\cdot10^{16}$ cm or expressed in terms of bullet radii
$60r_b \ \times \ 8r_b$. The ambient density is $\rho_a=1000 \ 
cm^{-3}$ and the ambient temperature $T_a=10^4 K$. It is assumed that
initially the bullet is in pressure equilibrium with the ambient
material.

The run was carried out in cylindrical symmetry with the x-axis
being the symmetry axis. We use an adaptive grid with 4 levels
of refinement. This provided a resolution of 128 cells per cloud
radius or an equivalent resolution of $1024 \ \times \ 7680$
cells. Although the above resolution is sufficient to place the
simulation in the converged regime in the adiabatic case
\citep*{KMC}, in the cooling case one has to be more precise in
one's definition of convergence. Since, as it will be discussed
below, the key process in the system is bullet fragmentation via
instabilities at the upstream surface, our criterion of
convergence was the constancy of the initial fragmentation
spectrum. In this sense the above resolution ensured that the
instability wavelength, and therefore the number of fragments into
which the bullet breaks up, does not change with increasing
resolution. On the other hand, it should be mentioned that such
resolution might not be sufficient to track the contraction of
each fragment due to radiative cooling to its smallest size
defined by the equilibrium between cooling and external heating.
Finally, outflow boundary conditions were prescribed on all
boundaries.

For the simulation described above the cooling parameter was $\psi_b =
2.8\cdot10^{-5}$. The hydrodynamic timescale, i.e. the bullet crushing
time, was $t_{hydro} \ = \ 2.54\cdot10^9$ s $\approx \ 80.5$ yrs., the
cooling timescale was $t_{cool} \ = \ 70.9\cdot10^3$ s $\approx \ 
19.7$ hours. The simulation total run time was $9.49\cdot10^{9}$ s
$\approx 251.1$ yrs\footnote{Note that the simulation run time is
  larger than the age of the $\eta$ Carinae strings ($\approx 150$
  yrs.). However, since it is the dimensionless numbers that determine
  the system evolution, the results can be appropriately scaled to
  $\eta$ Carinae conditions. We discuss that in further detail in
  Section 3.2.}. Thus the evolution of the system was in a regime
strongly dominated by cooling.

\section{RESULTS}

\subsection{Overall Morphology and Evolution of the System}

When the interaction of an inhomogeneity (bullet or clump) proceeds in
the adiabatic regime the dominant process defining the evolution of
the system is the lateral clump re-expansion \citep{KMC,Pol}. The
internal shock compresses the clump and once such compression is
completed the re-expansion begins into regions of lowest total
pressure, i.e. in the lateral direction. Such re-expansion, acting in
combination with instabilities at the upstream clump surface, is
responsible for the destruction of the clump. When more than one clump
is present, lateral re-expansion also drives interclump interactions
via merging and formation of larger structures that subsequently alter
the global flow \citep{Pol}.

The fundamental distinction between the radiatively cooled
inhomogeneous systems, including individual bullets, and the adiabatic
ones is the minimal role of lateral re-expansion. Instead, the
dominant process is the formation of instabilities at the upstream
surface of the bullet with a wavelength significantly smaller than the
one observed in the adiabatic case. As the bullet begins to drive
through the ambient medium, hydrodynamic instabilities
(Richtmeyer-Meshkov, Rayleigh-Taylor) produce the initial instability
seed. The resulting density variations quickly trigger the onset of
thermal instabilities which thereafter become the dominant process
responsible for bullet fragmentation. In our simulation such
instabilities started developing with a wavelength of about 8 - 10\%
of the initial bullet radius or $2.4\cdot10^{14} - 3.0\cdot10^{14}$ cm
$\approx \ 16 - 20$ a.u.  This initial scale is crucial for subsequent
evolution since it determines not only the continuing process of
bullet fragmentation but also the structure of the downstream flow. As
it was mentioned above, we did not observe any significant changes in
the fragmentation spectrum with changing grid resolution, therefore we
believe that we observed the true fragmentation spectrum corresponding
to the given flow conditions. Of course this conclusion may be altered
with fully 3-D simulations. The question of predicting the properties
of the initial fragmentation spectrum for the given characteristics of
the system is important for the understanding of the physics of
radiative hypersonic inhomogeneous systems, however it is outside the
scope of the current paper and it will be investigated in the
subsequent work.

Since the wavelength of the initial fragmentation spectrum is
approximately constant along the bullet radius, the fragments produced
by the instability are of different mass with the most massive ones
being closest to the symmetry axis and the outermost ones being the
lightest. As a result the fragments are ``peeled off'' from the bullet
one by one starting with the outermost ones in radius. Each
fragmentation event results in the formation of a distinct
``ring-like'' feature in the bullet wake. The formation of rings is a
consequence of the axisymmetry of these simulations. In 3-D it is
likely that the rings would themselves fragment \citep{Klein,Robey}.

One noteworthy point is that the evolution of the bullet and its
fragments suggests the presence of steady ``mass loading'' of the
downstream flow via the hydrodynamic ablation. Mass loading has been
claimed to be an important process in all clumpy hydrodynamic systems
\citep{HD86,HD88}. The extent and nature of such mass loading will be
investigated in the subsequent work.

\begin{figure}
\epsscale{1.0}
\plotone{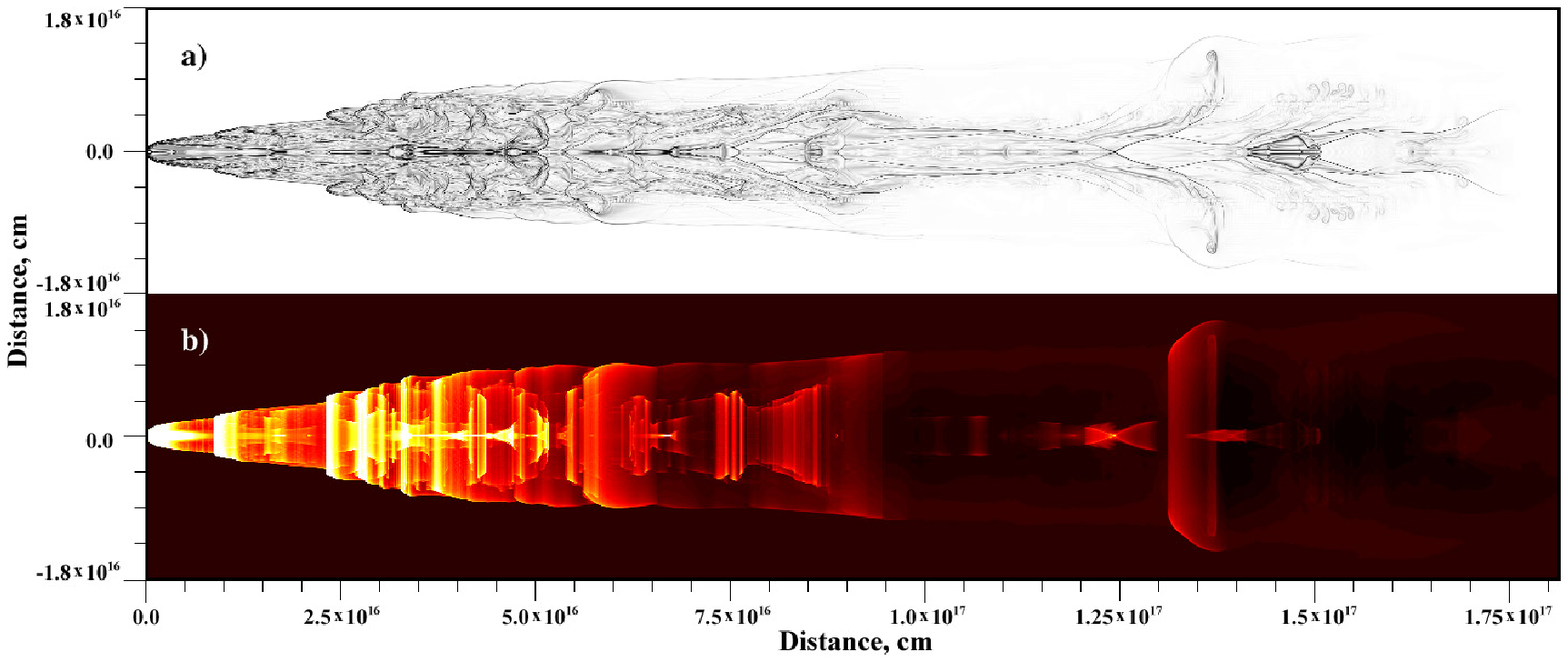}
\caption{a) Synthetic Schlieren image of the computational domain at time 251.1 yrs.
  Shown is the gradient of the density logarithm. b) Synthetic
  observation image of the computational domain for the same time as
  in a) (see text for the detailed description). Note the periodic
  ring-like structures in the domain resulting from individual
  fragmentation episodes.
\label{LastFrame}}
\end{figure}

Figure~\ref{LastFrame} shows the computational domain at time $t \ = \ 
251.1$ yrs. In Figure~\ref{LastFrame}a the synthetic Schlieren image
(gradient of the density logarithm) is shown illustrating the shock
and vortex sheet structure in the flow. Figure~\ref{LastFrame}b shows
the synthetic observation image of the computational domain. Since our
simulation did not track the full ionization dynamics of the flow, the
image represents the total radiative energy losses summed along each
ray. The 2D distribution of the state vector obtained in the
simulation was extended using cylindrical symmetry to a $2048 \times
2048 \times 7680$ cells uniform grid 3D data cube. Thus the synthetic
observation image represents the 2D projected distribution of the
logarithm of the emissivity $I$ integrated in the z-direction
according to the formula \beq
I_{ij}=\sum_kn_{ijk}^2\Lambda_{ijk}(T_{ijk}),
\label{emissivity}
\eeq
where $i,\ j,$ and $k$ are the cell indices in the x-, y-, and
z-direction respectively, and the cooling rate $\Lambda(T)$ here, as
well as in the simulation, was determined based on the cooling curve
described by \citet{Dalgarno}.

There are several ring-like structures in Figure~\ref{LastFrame} in
the bullet wake.  These result from the fragmentation events mentioned
above. The most prominent ring occurs at the distance of $\approx
1.38\cdot10^{17}$ cm from the bullet head (ring 1) with the width of
$\approx 2.94\cdot10^{16}$ cm $\approx 1965$ a.u. which is the widest
part of the bullet wake. In Figure~\ref{LastFrame}b several other
bright rings, resulting from fragmentation events, are visible closer
to the bullet head. They gradually increase in radius further
downstream with the radius of the largest ring (ring 2), located at
the distance of $\approx 5.8\cdot10^{16}$ cm from the bullet head,
being $\approx 1.82\cdot10^{16}$ cm $\approx 1216$ a.u. Note that the
width of the bullet wake stays practically constant, increasing only
slightly, for a distance of $\approx 8.0\cdot10^{16}$ cm $\approx
5350$ a.u. from ring 2 to ring 1. Thus in our simulations the
length-to-width ratio of the bullet wake is $6.1-9.9$ and the bullet
wake width is $4-6.5$ times larger than the width of String 5 while
they share the same length. Note also that the width of the bullet
head is only $\approx 1.05\cdot10^{15}$ cm $\approx 70$ a.u. We
followed the simulation beyond the time $t=251.1$ yrs, shown in
Figures~\ref{LastFrame} and \ref{Velocity}, up to the time $t=300.8$
yrs. and we did not observe any appreciable changes in the
aforementioned values of the length-to-width ratio.


The second process that determines the structure of the flow in the
wake is gas re-expansion into the cavity excavated by the bullet. The
highest temperature reached by gas in the system is at the tip of the
central bullet fragment and is about $3.5\cdot10^5$ K. Gas passing
through the bullet bowshock cools down very rapidly, weakening the
bowshock and making it more oblique, which in its turn prevents
further heating of the ambient gas. As it can be seen in
Figure~\ref{LastFrame}a, the bowshock practically disappears half-way
downstream from the bullet head.  Beyond the turbulent region
immediately behind the bullet head and in between the stripped bullet
fragments, the gas tends to re-expand essentially at the sound speed
of the ambient gas and fill the cavity. Such re-expansion causes gas
to rebound on the axis resulting in a reflected shock. The effect of
this shock can be seen in Figure~\ref{LastFrame}a around the symmetry
axis as a complex shock region that is narrower than the bowshock.

Note, that in Figure~\ref{LastFrame}b the surface brightness of the
system increases toward the bullet head. This differs from the surface
brightness behaviour observed in the $\eta$ Carinae strings. As was
noted by \citet{Weis99} the string surface brightness decreases toward
the head so that the true string tip might not even be detected.
Recall, however, that Figure~\ref{LastFrame}b is not a true synthetic
observation image in a certain line but rather a projected map of
total radiative energy losses in the system. In addition illumination
by stellar photons is not included in our calculation. Thus in the
simulation the bullet head and its vicinity, where the density and
temperature are highest so that $n^2\Lambda(T)$ are large, appear
brightest. In reality gas in the bullet head and near it may be too
hot to be visible optically as was suggested by \citet{Redman}.  In
our further work we plan to include tracking of full ionization
dynamics that will allow us to produce true synthetic observation
images in a particular emission line or a combination of lines.

\subsection{Velocity Distribution and Hubble-type Flow}

\begin{figure}
\epsscale{1.0}
\plottwo{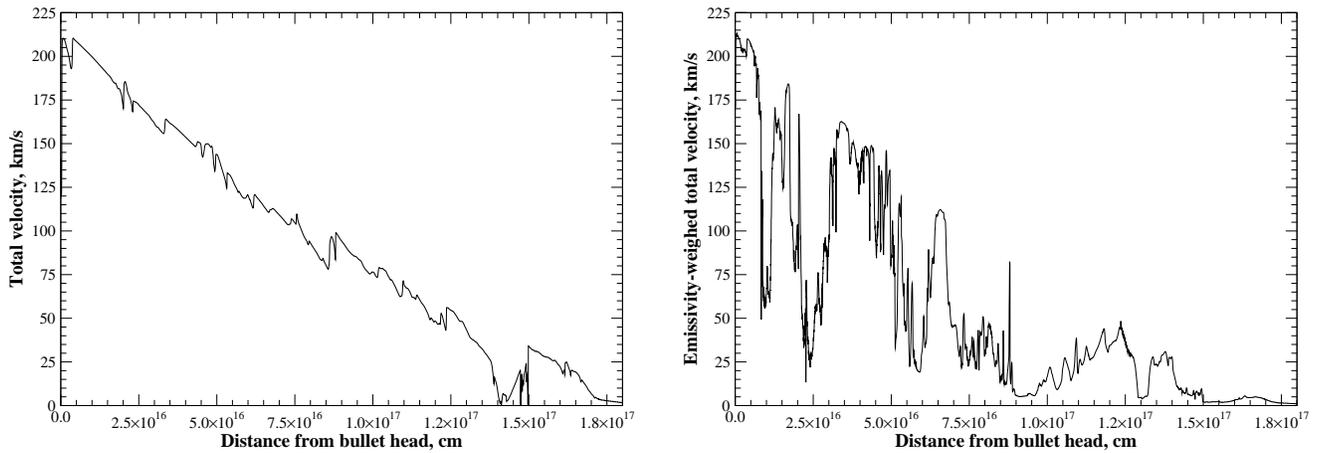}{f2b.eps}
\caption{\emph{Left:} distribution of total velocity along the symmetry axis of
  the bullet at the time 251.1 yrs. (same time in the simulation as
  the one shown in Figure~\ref{LastFrame}). \emph{Right:} distribution
  of the emissivity-weighed total velocity in the system at the same
  time (see text for the detailed description).
\label{Velocity}}
\end{figure}

Another key property of the strings is the presence of Hubble-type
flows in the bullet wake. The left panel of Figure~\ref{Velocity}
shows the distribution of the total velocity $v_{tot}(x)$ along the
symmetry axis of the bullet/wake as a function of distance from the
bullet head. The linear velocity decrease from the maximum value of
$210$ \kms from head to base is clear aside from some minor
fluctuations arising due to the unsteadiness of the downstream
flow\footnote{It should be noted that we observed similar behaviour of
  the total velocity distribution in other systems with the cooling
  parameter up to $\psi_b \approx 0.25$. For systems approaching the
  adiabatic regime with $\psi > 1.0$ the velocity distribution takes a
  more complicated form.}. Note that very little deceleration of the
bullet head has occurred. Given that the initial bullet velocity is
235 \kms we see that after 251.1 yrs., in which the bullet material
traveled the distance of $1.8\cdot10^{17}$ cm $\approx 0.058$ pc, the
bullet material lost only $\approx$ 10\% of its original velocity.
That is due to the fact that the central fragment retains most of the
original bullet mass while it has a rather small cross-section due to
the cooling-induced contraction.

\citet{Weis99} quote a velocity slope in the $\eta$ Carinae strings
ranging from 2590 \kms pc$^{-1}$ (String 5) to 3420 \kms pc$^{-1}$
(String 2). In our simulation we find a velocity slope of 3600 \kms
pc$^{-1}$ which is about 5\% - 40\% higher. Note, that the length of
String 5 is practically equal to the distance over which we allowed
the bullet to propagate, however String 2 is about 25\% shorter
\citep{Weis99}.

One should be cautious, however, in directly comparing the total
velocity distribution shown in the left panel of Figure~\ref{Velocity}
to the observationally determined velocity distributions in the
strings of $\eta$ Carinae (cf. Figure 6 in \citep{Weis99}).
Quantities, more closely resembling the ones that are obtained
observationally, should be employed in order to make the comparison
between the numerical results and observations relevant. For example,
one can create emissivity-weighed total velocity $v_{emis}(x)$ maps of
the flow. Distribution of such a quantity along the x-axis is shown in
the right panel of Figure~\ref{Velocity}. $v_{emis}(x)$ was obtained
according to the formula
\beq
v_{tot}(x)=\frac{\sum_jn^2_j\Lambda_jv_j}{\sum_jn^2_j\Lambda_j}\approx
\frac{\int_{y_{min}}^{y_{max}}n^2(x,y)\Lambda(x,y)v_{tot}(x,y)dy}
{\int_{y_{min}}^{y_{max}}n^2(x,y)\Lambda(x,y)dy},
\label{vtotemis}
\eeq
where $i$ and $j$ are the cell indices in the x- and
y-directions, $v_j$ is the total velocity in the cell $j$, and in the
summation in the numerator we included only the cells with non-zero
total velocity. The distribution of this quantity is significantly
more noisy than the total velocity cross-cut.  However, the local
maxima of $v_{emis}(x)$ roughly tend to fall on the same line as that
in the previous figure giving some indication of the presence of the
Hubble-type flow.

The distance traveled by the bullet in our simulation is practically
equal to the length of String 5 and is factors of $1.5 - 3.0$ shorter
than the rest of the strings (except for String 2) \citep{Weis99}. If
we assume that the true length of strings is not their observed length
but rather the distance of the tip from the star, than the $\eta$
Carinae strings are the factors of $3.2 - 5.56$ longer \citep{Weis99}
than the computational domain in our simulation. However, the
kinematic age of the bullet in our simulation is about 67\% larger
than that of the $\eta$ Carinae strings: 251.1 years vs. 150 years.
This indicates that in our simulation the bullet velocity was
approximately a factor $2.5 - 5.0$ (or a factor $5.35 - 9.3$ if we
assume that the string base is located near the central star) smaller
than in the case of $\eta$ Carinae strings, i.e. the bullet velocity
should be $590 - 1175$ \kms (or $1260 - 2185$ \kms) with respect to
the central star. An increased bullet velocity would decrease the
width of the bullet wake and increase the length-to-width ratio making
a better match between our simulations and the strings of $\eta$
Carinae. However, a significant increase in velocity, and therefore
Mach number, results in much higher temperatures in the bullet wake
and the bow shock. That, in turn, increases the cooling parameter
$\psi_b$ eventually sending the system into the adiabatic regime which
no longer exhibits the high length-to-width ratio of the wake and the
Hubble-type downstream flow\footnote{We observed that effect in a
  simulation of Mach 200 bullet propagation. The cooling parameter for
  that case was $\psi_b=3.7$. We do not discuss this simulation in
  this paper however its animation can be accessed at the web site
  mentioned in the acknowledgment section.}. The easiest way to
reconcile the need for the significantly higher bullet velocity,
required in order to obtain the correct kinematic age, with the need
for the relatively low post shock temperatures, required to decrease
the bullet wake width, is to embed the bullet in the ambient flow that
itself moves with significant speed relative to the central star.
There is some observational evidence for the presence of such high
velocity ambient flow \citep*{Weis01}. For example, string 1 may be
embedded in the material moving with velocity $\approx 500-600$ \kms
with respect to the central star, whereas the maximum velocity of the
string material is $\approx 995$ \kms \citep{Weis99}.

\section{CONCLUSIONS}

In this paper we have presented a numerical study of a hypersonic
radiative cosmic bullet in an attempt to model the strings observed in
the $\eta$ Carinae Homunculus nebula. Our simulations follow from the
discussion by \citet{Redman}.

Our principal conclusions are as follows: (1) hypersonic radiative
bullets are capable of producing structures with high length-to-width
ratios (between 6 and 10 for our study); (2) the dominant process
responsible for bullet destruction is instability formation at the
bullet upstream interface leading to separate fragmentation episodes;
these result in the formation of periodic ``ring-like'' structures in
the bullet wake; (3) the simulations do show the presence of
Hubble-type flows along the axis in the bullet wake in terms of a
linear decrease of the total velocity downstream from the bullet head.
How these flows appear observationally remains an open question.

Thus we conclude that the bullet model appears as a good candidate for
the strings of $\eta$ Carinae, though further work is needed.  If this
model finds continued success then theorists will confront the issue
of how such high velocity bullets are generated, i.e. at the star or
at larger distances. Confirming the existence of such bullets may be
helpful in determining the nature of the processes, occurring within
the star, which were part of its various eruptions.

The astrophysical applicability of the model, discussed in this
paper, might extend beyond the $\eta$ Carinae strings. One of the
most spectacular examples of nebular systems with long thin
structures is the protoplanetary nebula CRL 618 and, in
particular, its shocked lobes. The most prominent features of the
lobes are the periodic rings similar to the ones present in our
simulation. As it was discussed, such rings arise naturally as a
consequence of the bullet fragmentation events. There is also some
evidence for the velocity increase in the lobes from the base to
the tip \citep*{Sanchez}. Other examples include, but are not
limited to, other planetary nebulae, e.g. strings in NGC 6543
\citep{Weis99}, HH objects, etc.

Further developments of the model should include the study of the
details of the formation of the initial fragmentation spectrum,
investigation of the importance of mass-loading due to the ablation of
the bullet head and bullet fragments, and inclusion of more realistic
description of ionization dynamics and radiative cooling in the
system. The latter will allow us to produce more realistic synthetic
observation images and synthetic spectra for better comparison with
observational data.

\acknowledgements

This work was supported in part by the NSF grant AST-9702484, NASA
grant NAG5-8428 and the Laboratory for Laser Energetics under DOE
sponsorship.

The most recent results and animations of the numerical experiment,
described above as well as the ones not mentioned in the current
paper, can be found at \url{www.pas.rochester.edu$/^{\sim}$wma}.


\clearpage

\end{document}